%% file: main.tex
\PassOptionsToPackage{table,xcdraw}{xcolor}
\documentclass[sigconf,language=english]{acmart}

\copyrightyear{2025}
\acmYear{2025}
\setcopyright{cc}
\setcctype{by}
\acmConference[WiSec 2025]{18th ACM Conference on Security and Privacy in Wireless and Mobile Networks}{June 30-July 3, 2025}{Arlington, VA, USA}
\acmBooktitle{18th ACM Conference on Security and Privacy in Wireless and Mobile Networks (WiSec 2025), June 30-July 3, 2025, Arlington, VA, USA}
\acmDOI{10.1145/3734477.3734722}
\acmISBN{979-8-4007-1530-3/2025/06}

\usepackage{svg}

\usepackage{enumitem}
\usepackage{longtable,booktabs}
\usepackage{array}
\usepackage{titlesec}
\IfFileExists{footnotehyper.sty}{\usepackage{footnotehyper}}{\usepackage{footnote}}
\makesavenoteenv{longtable}
\newcommand{\mh}[1]{\textcolor{teal}{\textbf{MH:} #1}}
\newcommand{\todo}[1]{{\color{red}{\textbf{TODO:} #1}}}

\newcommand{\cut}[1]{{\color{gray}{#1}}}
\usepackage{multirow}
\usepackage{tablefootnote}
\usepackage{graphicx}
\usepackage{acro}
\usepackage{siunitx}
\usepackage{tablefootnote}
\usepackage{arydshln}
\acsetup{patch/longtable=false}

 \renewcommand{\mh}[1]{}
 \renewcommand{\todo}[1]{}

 \renewcommand{\cut}[1]{}

\input{acronyms}

\begin{document}

\title{Assessing the Latency of Network Layer Security in 5G Networks}

\author{Sotiris Michaelides}
\orcid{0009-0003-6020-3934}
\affiliation{%
  \institution{RWTH Aachen University}
  \city{Aachen}
  \country{Germany}
}
\email{michaelides@spice.rwth-aachen.de}

\author{Jonathan Mucke}
\orcid{0009-0008-7506-948X}
\affiliation{%
  \institution{RWTH Aachen University}
  \city{Aachen}
  \country{Germany}
}
\email{jonathan.mucke@rwth-aachen.de}

\author{Martin Henze}
\orcid{0000-0001-8717-2523}
\affiliation{%
  \institution{RWTH Aachen University}
  \city{Aachen}
  \country{Germany}
}
\additionalaffiliation{%
  \institution{Fraunhofer FKIE}
  \city{Wachtberg}
  \country{Germany}  
}
\email{henze@spice.rwth-aachen.de}

\begin{abstract}
In contrast to its predecessors, 5G supports a wide range of commercial, industrial, and critical infrastructure scenarios. 
One key feature of 5G, ultra-reliable low latency communication, is particularly appealing to such scenarios for its real-time capabilities.
However, 5G's enhanced security, mostly realized through \emph{optional} security controls, imposes additional overhead on the network performance, potentially hindering its real-time capabilities. 
To better assess this impact and guide operators in choosing between different options, we measure the latency overhead of IPsec when applied over the N3 and the service-based interfaces to protect user and control plane data, respectively. 
Furthermore, we evaluate whether WireGuard constitutes an alternative to reduce this overhead.
Our findings show that IPsec, if configured correctly, has minimal latency impact and thus is a prime candidate to secure real-time critical scenarios.

\end{abstract}

\ccsdesc[500]{Security and privacy~Security protocols}
\ccsdesc[500]{Security and privacy~Mobile and wireless security}
\ccsdesc[300]{Networks~Mobile networks}

\keywords{5G; Network Layer Security; Latency; IPsec; WireGuard; TLS}

\maketitle

\section{Introduction}
\label{sec:intro}

Up until the Fourth Generation (4G) of mobile networks, the primary focus was on enhancing mobile broadband for commercial use, prioritizing bandwidth \cite{10.5555/2124884.2124897}. 
However, the advent of Fifth Generation (5G) networks marked a significant shift towards addressing both commercial \emph{and} industrial deployments.
This transition targeted industrial applications and critical infrastructure, meeting the growing demand for robust, low-latency communication~\cite{michaelides2025industry5G,ts22261}.

To this end, 5G supports \ac{urllc}, crucial for real-time applications, making it an ideal backbone for critical infrastructure such as industrial control systems and healthcare.
In these sectors, even minor latency shifts can disrupt production lines or create safety hazards in real-time data transmission scenarios, such as remote surgeries.

At the same time, 5G's enhanced security over its predecessors resonates extremely well with the strict security requirements of such sectors faced by growing security threats~\cite{stellios2018survey}. %
Concretely, 5G implements advanced security protocols and controls addressing previous vulnerabilities, such as the lack of \ac{up} integrity protection in 4G \cite{8835335}. 
These controls and protocols are essential for ensuring security guarantees, such as protection against tampering and eavesdropping attacks.
However, the utilization of many of these is optional and left to the discretion of network operators.

A notable example is the use of network layer security to ensure these essential security guarantees. 
The 3GPP 5G specification suggests implementing IPsec on most 5G interfaces, including those within the \ac{5gc}, which manages critical functions such as authentication, billing, and security. 
Without proper protection, these interfaces remain vulnerable to data tampering, privacy breaches, or service outages. 
While the specification mandates their protection, utilizing security protocols can be avoided if alternative measures, such as physical security, are used~\cite{ts33501}.
Furthermore, even if network operators principally decide to apply such optional security controls, they have to choose between different configuration options, e.g., ciphers, authentication options, and operation modes.%

This decision is further complicated as, besides tremendous security benefits, these optional controls increase latency, contradicting efforts to achieve reliable communication with minimal latency. 
Still, with the rise of distributed and cloud-based \ac{5gc} deployments~\cite{gsma2024}, where physical protection becomes impractical, network operators must consider enabling these controls to improve security. 
Consequently, they require a profound understanding of how these controls impact 5G's ability to support time-critical applications and which configurations are most latency-friendly. %

This paper provides this much-needed understanding by empirically evaluating the latency overhead of network layer security in 5G. Specifically, we analyze the latency impact of \acs{ipsec} in various configurations on both the \acf{up} and the \ac{cp} of the \ac{5gc} in a simulated setup. 
Furthermore, we assess the overhead of WireGuard, a modern security protocol often proposed as a promising alternative~\cite{tlscore,haga20205g}. In detail, our contributions are:

\begin{enumerate}[topsep=3pt,leftmargin=14pt]

    \item By orchestrating and extending existing open-source components, we facilitate the first open 5G testbed that implements IPsec and WireGuard \cite{anonymous2025assessing}.

    \item By measuring the overhead of IPsec for UP data transmission and CP communication within the 5GC, we identify latency-friendly IPsec configurations that add an overhead of just \SI{55}{\micro\second} for UP and between \SI{300}{\micro\second}-\SI{600}{\micro\second} for CP.

    \item By assessing WireGuard, we find that while it shows no advantage in w.r.t latency, it is still a fast and more resource-friendly alternative compared to IPsec.

\end{enumerate}

\section{Background on 5G Systems and Their Security}

\begin{figure}[] 
    \centering
    \includegraphics[width=0.475\textwidth]{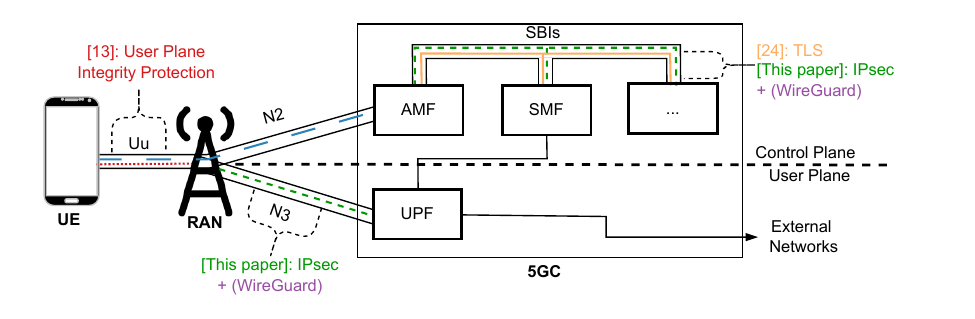} 
    \vspace{-15pt}
    \caption{In a 5G network, separated into control and user plane, various security controls apply to different planes and interfaces: \textcolor[HTML]{2B83BA}{Control Plane Security}, \textcolor[HTML]{d7191c}{User Plane Security}, \textcolor[HTML]{009900}{IPsec}, and \textcolor[HTML]{fdae61}{TLS}.
    In this work, we study the latency impact of \textcolor[HTML]{009900}{IPsec} and \textcolor[HTML]{984EA3}{WireGuard} to secure the \ac{up} of 5G and the \ac{cp} of the \ac{5gc}.
    }
    \label{fig:5G}
\end{figure}

A 5G system comprises the \ac{ue}, \ac{ran}, and \ac{5gc} (cf.\ Figure~\ref{fig:5G}). 
The \ac{ue} is the end-user device equipped with authentication credentials while the \ac{ran} manages radio resources to provide wireless connectivity to the \ac{ue} via the Uu interface. 
The \ac{5gc}, linked to the \ac{ran}, interconnects \acp{nf} controlling network operations and managing connections~\cite[§4.2]{ts23501}.
A 5G system is logically divided into two planes: the \acf{cp} handling control functions such as authentication and session management, and the \acf{up} transmitting user data~\cite[§4.3]{ts23214}. 
This separation is most evident in the \ac{5gc}, where control tasks are handled by \ac{cp} \acp{nf} such as the \ac{amf}, \ac{smf}, and \ac{pcf}, while the \ac{upf} in the \ac{up} routes user data to external networks.
\ac{cp} \acp{nf} communicate with each other over \acp{sbi}, while \ac{cp} and \ac{up} data are transmitted from the \ac{ran} to the \ac{amf} and \ac{upf} via the N2 and N3 interfaces, respectively.

\subsection{Security Controls in 5G}

To secure data over 5G, 3GPP defines several security controls, which inevitably affect network performance, especially latency.

\textbf{CP and UP Security} The specification mandates security controls for both planes, comprising three encryption and three integrity protection schemes based on AES, SNOW, and ZUC~\cite{bicmac}. 
However, operators are only required to enable integrity protection for \ac{cp} data, while the other security controls remain optional. 
While both planes use the same schemes, their termination points differ: \ac{cp} security is established between the \ac{ue} and \ac{amf}, whereas \ac{up} security terminates at the \ac{ran}. 
Thus, control data from the UE is always integrity-protected over Uu and N2.
On the other hand, even with UP security enabled, user data is only protected over Uu, remaining unprotected over N3 unless higher-layer end-to-end security is applied. 
This distinction is critical for low-latency communication, as securing user data across 5G requires \emph{both} UP and N3 security, introducing two latency overheads (cf. Figure~\ref{fig:5G}).

 \textbf{ N3 Security.} Securing \ac{up} data over the N3 interface (i.e., between \ac{ran} and \ac{upf}), requires the utilization of \ac{ipsec}, configured with specific parameters such as cryptographic algorithms and corresponding key lengths, as mandated by 3GPP. 
However, operators may optionally choose not to implement these security measures if the \ac{ran} is placed in a ``\emph{secure environment}''~\cite[§9.3]{ts33501}.

\textbf{SBI Security.} Similar to the N3 interface, SBIs within the 5GC must be secured unless deemed ``\emph{trusted (e.g., physically protected)}''~\cite[§13.1]{ts33501}. 
To secure SBIs, the specification mandates the support   of \ac{tls}, but also notes NDS/IP (3GPP's specification for IPsec configuration) as an alternative option.

\subsection{Security Protocols} 
\label{label:sec_prot}
\textbf{5G Standardized Protocols.} \textit{\ac{ipsec}} is a suite of protocols that provides security between two hosts at the network layer. 
Initially, the \textit{\ac{ike}} protocol handles mutual authentication and establishes security associations by exchanging certificates, cryptographic challenges and key material, usually in two round trips. %
The \textit{\ac{esp}} protocol then establishes the secure tunnel using the parameters negotiated in \ac{ike}. %
\ac{esp} can operate in transport mode (encrypting the payload) or tunnel mode (encrypting the entire packet), with tunnel providing more security by hiding topology information \cite[§9.2]{ts33501}. 
While both protocols introduce additional overhead, \ac{ike} has a greater impact, especially in scenarios that require multiple authentication events, as it is often the case within the \ac{5gc}.

\textit{\ac{tls}} in contrast realizes end-to-end security at the transport layer~\cite{tlscore}. 
Similar to \ac{ipsec}, it uses an initial handshake to exchange certificates and key material for authentication. 
Afterwards, a secure channel is established. %
The current standards, TLS 1.2 and TLS 1.3, differ in latency impact; TLS 1.3 requires only one round trip for the handshake compared to two in TLS 1.2. %

Both \ac{ipsec} and \ac{tls} are established security protocols that offer flexibility by supporting various cryptographic algorithms and key lengths as well as authentication using either \ac{psk} or certificates.
In this work, we identify 12 mandatory configurations for \ac{ipsec}, while related work identifies 14 for \ac{tls}~\cite{tlscore}.

\textbf{WireGuard as Alternative.}
\textit{WireGuard} is a relatively new protocol (proposed in 2017), that challenges both \ac{ipsec} and \ac{tls} w.r.t.\ latency and bandwidth~\cite{Donenfeld2017WireGuardNG}.
Although not part of the 5G standards, it is discussed as an alternative to improve latency in 5G~\cite{tlscore,haga20205g}. 
Similar to IPsec, it realizes security at the network layer. 
It utilizes a simple, small-size, one-round-trip handshake for authentication.
Then, a modern  Authenticated Encryption with Associated Data (AEAD) algorithm, CHACHA20\_POLY1305, is used to establish the secure channel. 
AEAD algorithms are faster than traditional schemes as they perform encryption and authentication simultaneously. 
In contrast to \ac{ipsec} and \ac{tls}, WireGuard intentionally only supports one configuration with \acp{psk} and 256-bit keys.

\subsection{Related Work}\label{sec:related-work}

While 5G is well-established, the impact of its (optional) security controls, especially w.r.t latency, has received limited attention. Heijligenberg et al.\cite{bicmac} evaluate the impact of UP integrity protection (\textcolor[HTML]{d7191c}{red} in Figure\,\ref{fig:5G}), showing noticeable effects across all schemes, even in low-latency configurations. Zeidler et al.\cite{tlscore} assess the impact of \ac{tls} in the \ac{5gc} (\textcolor[HTML]{fdae61}{orange} in Figure\,\ref{fig:5G}), finding minimal impact on \ac{ue} registration and PDU establishment in a running network but significant overhead on a freshly started one. 
Haga et al. evaluate WireGuard and OpenVPN (TLS) for slice isolation, finding that WireGuard outperformed OpenVPN across all metrics\cite{haga20205g}.

Without focusing on 5G, several works compare the performance of security protocols. Kotuliak et al. compare \ac{ipsec} and \ac{tls} for interconnected IP multimedia subsystems, finding comparable performance, with a slight advantage for \ac{ipsec}. Similarly, Dekker et al.~\cite{Dekker2020PerformanceCO} compare WireGuard, Strongswan (IPsec), and OpenVPN in 1 Gbit/s environments, showing Strongswan ciphers outperforming the others in terms of latency. 
In contrast, Donenfeld~\cite{Donenfeld2017WireGuardNG} reports that both WireGuard and \ac{ipsec} add less latency than TLS, with WireGuard performing best. However, the WireGuard website notes that these results are \textit{“old, crusty, and not super well conducted”}~\cite{WireGuardPerformance}.

\textbf{Novelty of Our Work:}
As highlighted in Figure~\ref{fig:5G}, our work significantly extends prior research in the following manner. 
First, while previous work on UP data protection~\cite{bicmac} focused on the UE-RAN link, we expand this by evaluating 3GPP standardized security controls over N3, offering a comprehensive assessment of UP data protection across 5G. 
Second, we complete the evaluation of 3GPP-standardized security protocols within the 5GC.
 While previous research has analyzed TLS~\cite{tlscore}, we extend this by evaluating IPsec and reproducing some of the best-performing TLS configurations for comparison. 
Finally, we assess WireGuard as a potential alternative for protection over N3 and within the 5GC, providing an up-to-date comparison of the three protocols.

\section{Methodology}

To provide a profound understanding of the latency impact of 5G's network layer security, we perform latency measurements over the N3 interface during UP data transmission and over the SBI during the UE attachment to the network. 
In the following, we first discuss our methodology for extracting mandatory IPsec configurations from 3GPP standards (§\ref{label:config}), before we describe the containerized deployment used to conduct our measurements (§\ref{sec:experimental-setup}).

\subsection{Selection of IPsec Configurations} 
\label{label:config}

IPsec can be configured in various ways, involving different modes of operation, ciphers, and authentication algorithms. To ensure baseline compatibility among 5G components implementing IPsec, we adhere to the 3GPP specification TS 33.501, which outlines the security architecture and procedures for 5G systems. 
According to TS 33.501 \cite[§9.3,13.1.0]{ts33501}, IPsec configurations must comply with TS 33.210\cite{ts33210} and TS 33.310 \cite{ts33310}, applicable to both N3 and \acp{sbi}. For ESP configuration, the standards refer to RFC 8221 \cite{RFC8221}.

From these documents, we identify and test the mandatory-to-support configurations for both \ac{ike} and ESP to ensure alignment with compatibility and security standards in 5G deployments. 
We exclude mandatory-to-support configurations explicitly classified as not recommended, such as RSA signatures with PKCS\#1 v1.5 padding—expected to be prohibited by 2030~\cite[§6.1]{ts33310}—and the utilization of the Authentication Header, which is discouraged as ESP can provide encryption and authentication more efficiently \cite[§4]{RFC8221}.

\begin{table}[t]
    \centering
    \caption{Our experiments cover all IPsec configurations which are defined as mandatory to support in 5G.} %
    \vspace{-5pt}
    \small %
    \begin{tabular}{@{}>{\centering\arraybackslash}p{0.7cm} >{\centering\arraybackslash}p{1.7cm} >{\centering\arraybackslash}p{3cm} >{\centering\arraybackslash}p{1.8cm}@{}} %
        \toprule
        \textbf{} & \textbf{Type} & \textbf{Configuration} & \textbf{Document} \\ 
        \midrule
        \multirow{6}{*}{\rotatebox{90}{\textbf{IKE}}} & Encryption & AES128-GCM, ICV128  & TS 33.210 §5.4.2 \\ 
        & \cellcolor{gray!10}PRF & \cellcolor{gray!10}HMAC-SHA256  & \cellcolor{gray!10}TS 33.210 §5.4.2 \\ 
        & Integrity  & HMAC-SHA256-128  & TS 33.210 §5.4.2 \\  
        & \cellcolor{gray!10}Key exchange & \cellcolor{gray!10}DH Group 19  & \cellcolor{gray!10}TS 33.210 §5.4.2 \\ 
       & \multirow{2}{*}{Authentication} & ECDSA-SHA256$^\alpha$ & TS 33.310 §6.2.1 \\
        &  & SKMIC$^\beta$ & RFC 7296 §3.8 \\ 
        \midrule
        \multirow{7}{*}{\rotatebox{90}{\textbf{ESP}}} & \cellcolor{gray!10} & \cellcolor{gray!10}AES128-GCM, ICV128 & \cellcolor{gray!10}RFC 8221 §5 \\  
        & \multirow{-2}{*}{\cellcolor{gray!10} AEAD} & \cellcolor{gray!10}AES256-GCM, ICV128 & \cellcolor{gray!10}RFC 8221 §5 \\  
        \cdashline{2-4}
        & \multirow{3}{*}{Encryption} & AES128-CBC & RFC 8221 §5 \\  
        & & AES256-CBC & RFC 8221 §5 \\  
        & & NULL & RFC 8221 §5 \\  
        & \cellcolor{gray!10} & \cellcolor{gray!10}HMAC-SHA256-128 & \cellcolor{gray!10}RFC 8221 §6 \\  
        & \multirow{-2}{*}{\cellcolor{gray!10} Integrity} & \cellcolor{gray!10}GMAC-AES128$^\gamma$ & \cellcolor{gray!10}TS 33.210 §5.3.4 \\  
        \bottomrule
        \multicolumn{4}{l}{\scriptsize 
$^\alpha$Certificate Authentication \hspace{1em} 
$^\beta$ PSK Authentication \hspace{1em} 
$^\gamma$ Only with \textit{NULL} encryption} \\
    \end{tabular}
    \label{tab:ipsec-config}
\end{table}

We summarize the IPsec configurations used in our study in Table \ref{tab:ipsec-config}. 
For IKEv2, we identify one mandatory set of configurations that can be used with either certificates or \acp{psk} for authentication, resulting in two distinct test cases.
For the ESP configuration, either an AEAD algorithm may be used, or a combination of separate algorithms for encryption and integrity (cf. dashed line in Table\,\ref{tab:ipsec-config}). 
While integrity protection can be applied without encryption, the reverse (encryption without integrity protection) is prohibited by RFC 8221~\cite[§4]{RFC8221}.
Based on this, we identify a total of six distinct test cases for ESP. 
Thus, the combination of the two protocols results in twelve ($2$ \ac{ike} $\times$ $6$ \ac{esp}) mandatory-to-support IPsec configurations. 
Finally, we configure \ac{ipsec} in tunnel mode, as this setup provides an additional layer of security, specifically topology hiding by also encrypting the IP headers~\cite[§9.2]{ts33501}.

Besides IPsec, we note that WireGuard requires no configuration, as it intentionally supports only one configuration (cf.~§\ref{label:sec_prot}).
For our reproduction of TLS measurements over the SBI, we use TLS 1.2 and 1.3 with AES-GCM at 128-bit and 256-bit key lengths, as these have been shown to perform best w.r.t.\ latency \cite{tlscore}.

\subsection{Experimental Setup}
\label{sec:experimental-setup}

To measure the latency impact of these security protocol configurations, we set up a testbed based on a containerized deployment of open-source components, which we make publicly available ~\cite{anonymous2025assessing}.

\textbf{Testbed:} %
We rely on open-source 5G components widely used in academia: 
Open5GS~\cite{open5gs2024} for the \ac{5gc} and UERANSIM for the RAN and UE simulation~\cite{ueransim2024}. 
Each \ac{nf} as well as the \ac{ran} and \acp{ue} are deployed as separate Docker containers on the same physical host.
By realizing a controlled environment, where all components run on a single physical host, we ensure accurate measurements by minimizing external factors that could affect latency.

We add IPsec-secured communication by utilizing strongSwan~\cite{strongswan2024}. 
NFs communicating over the SBIs establish IPsec tunnels using pre-distributed certificates or PSKs. 
We also use IPsec to secure communication between the RAN and UPF over the N3 interface. 
In our deployment, strongSwan handles the traffic transparently, enabling its  easy replacement with other IPsec solutions.
Additionally, we modify Open5GS to fully support IPsec in trap mode (cf. §\ref{sub:control}), where the tunnel is established upon detecting traffic that matches the tunnel's policy.
In this mode, packets sent before the tunnel is established are dropped, causing slower SBI connection setups due to retransmissions. To mitigate this, we modify the NFs to send a dummy packet and synchronously wait for the tunnel to be established before proceeding.
To evaluate WireGuard, we use the WireGuard implementation in Linux kernel version 6.5.0-1027-oem.
For the replicated TLS measurements over the SBIs, we enable Open5GS's built-in TLS support in each NF's configuration file.

\textbf{Host Machine:} For our measurements, we deploy the containers on a computer running Ubuntu 22.04 LTS and equipped with an Intel Core i7-13700 (8 Performance and 8 Efficient cores, with a Performance-core base frequency of 2.10 GHz) as well as 64 GB of DDR5 RAM. 
We utilize AES acceleration through AES-NI provided by the CPU, as real-life 5G deployments are likely to operate on hardware that also supports AES acceleration.

\section{Latency Overhead of Tunneling Protocols}
\label{label:results}

We evaluate the latency overhead of security controls in 5G by measuring the time required for specific procedures compared to a baseline scenario (without security controls). 
We assess their impact on \ac{up} data transmission over the N3 interface (\S\ref{n3}) and the \ac{cp} data in the \ac{5gc} (\S\ref{sub:control}). 
For each experiment we report the mean over multiple runs with 99\% confidence intervals ensuring high accuracy in pinpointing the true average. 
Different configurations of \ac{ipsec} (and TLS) are presented as \textit{``Encryption''}\_\textit{``Integrity''}, or a single algorithm for AEAD schemes. 
We further investigate the CPU time of each protocol (\S\ref{sec:scalability}) and validate latency results with real UEs (\S\ref{sec:realues}).
Finally, we discuss our results with MNOs (\S\ref{sec:MNOs}).

\subsection{User Data Transmission over N3}
\label{n3}

To evaluate the latency overhead of IPsec and WireGuard on the N3 interface, we measure the round-trip time (RTT) between the UE and the host machine. The RTT reflects twice the transmission latency for \ac{up} data traveling from the UE to its destination. %

\textbf{Evaluation Method.}
To generate UP traffic, we use \texttt{Ping} within the UE container. 
\texttt{Ping} measures the RTT of a single packet between two hosts using ICMP. 
To minimize external factors that could influence our measurements, we ping the host machine, ensuring that traffic remains confined to the physical host.
As the RAN and the UPF establish and maintain the tunnel upon exchanging data for the first time—shortly after the first user packet—the impact of tunnel establishment (i.e., \ac{ike}) becomes negligible. 
Therefore, we only consider the six ESP configurations, and the single configuration of WireGuard, and start measuring after the tunnels are established.
We perform 20,000 repetitions for each configuration, divided into 10 sets of 2,000 repetitions. Additionally, we test different payload sizes: (i) 64 bytes (the default \texttt{Ping} packet size in Ubuntu) and (ii) 1,024 bytes (to amplify the impact of cryptographic operations). 

\begin{figure}[] 
    \centering
    \includegraphics[width=0.475\textwidth]{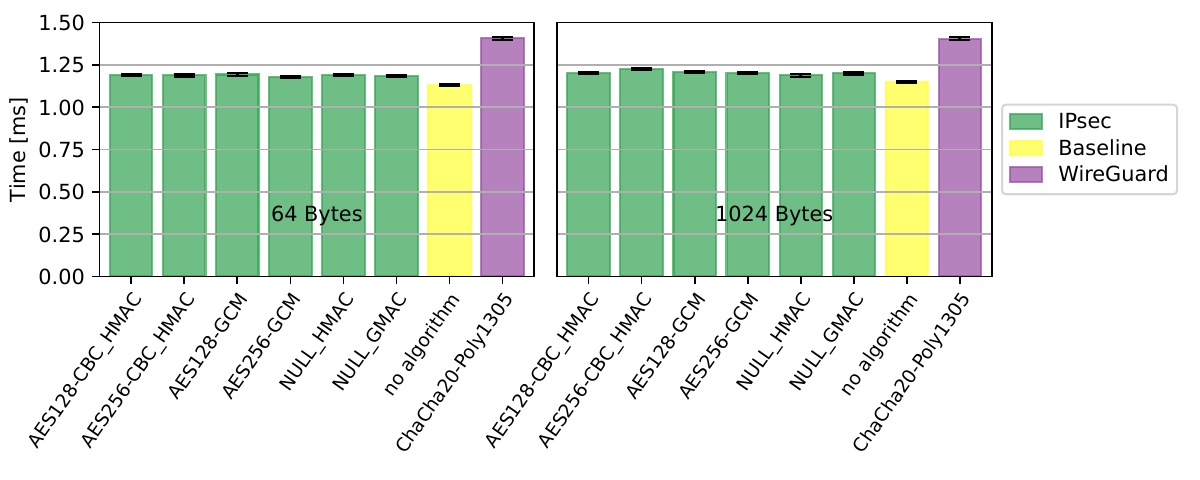} 
    \vspace{-20pt}
    \caption{The average RTT for user data transmission on the N3 interface behaves similarly for different payload sizes, with IPsec performing comparable to the insecure baseline while WireGuard shows a slightly worse performance.}
    \label{fig:n3}
\end{figure}

\textbf{Results.}
Our results in Figure~\ref{fig:n3} demonstrate a clear difference between the baseline scenario, \ac{ipsec}, and WireGuard. 
All six \ac{ipsec} ESP configurations show comparable performance, with an average latency overhead of \SI{55}{\micro\second} (for 64-byte payload) and a standard deviation of ±\SI{7}{\micro\second}, corresponding to a $\sim$5\% increase. 
In contrast, WireGuard introduces a larger overhead of \SI{260}{\micro\second}, making it 17.2\% slower than the fastest \ac{ipsec} configuration. 
For both, the overhead remains consistent as payload size increases. 
While the difference between them is noticeable, both protocols remain fast in absolute terms, with \ac{ipsec} being better suited for time-critical applications.   %

\subsection{Control Communication within the 5GC}
\label{sub:control}

To evaluate the impact of security protocols on the \ac{cp} of \ac{5gc}, we measure UE attachment time—from registration to transmission readiness.
During this process, NFs communicate with each other to fulfill tasks (e.g., authentication). By analyzing traffic in the 5GC, we identify 17 communication channels, including 14 previously found by Zeidler et al.~\cite{tlscore} and three new ones between the \ac{smf}/\ac{nrf}/\ac{pcf} and the Binding Support Function (BSF), which was recently introduced in Open5GS.

{\textbf{Evaluation Method.}}
Unlike measurements over N3, execution time in this scenario is primarily influenced by the number of connections that must be established within the 5GC CP. 
We analyze two scenarios: the cold scenario, where security associations between NFs are not established, and the warm scenario, where security associations are already in place, limiting overhead to encryption and integrity protection. 
These represent worst-case (cold) and best-case (warm) latency conditions.
While cold scenarios may seem unintuitive in (commercial) real-world deployments, where 5G systems are (almost) always operational, operators may deploy new NFs for scalability, recovery, and load balancing, requiring the re-establishment of many security associations \cite[§5.21.3.1]{ts23501}, leading to at least a partial cold scenario.
In the cold scenario, initial security setup adds significant overhead. 
However, of the 17 total channels needed for UE attachment, 8 are automatically established without traffic from the UE, as each NF contacts and registers with the NRF upon startup for NF discovery. 
Therefore, cold scenario measurements account for the time to establish 9 secure channels in addition to encryption and integrity protection.

We conduct measurements by booting up the 5GC and RAN, then sequentially deploying two UEs. The first UE represents the cold scenario, with no pre-established secure channels, while the second UE represents the warm scenario, leveraging channels created by the first. 
We capture traffic using Tshark and measure the time from the UE’s registration request to the PDU session establishment message sent by the SMF, signaling readiness for data transmission.
We measure all 12 IPsec configurations identified in §\ref{label:config} in tunnel mode and the single WireGuard configuration.
We also reproduce measurements for TLS 1.2 and 1.3 using AES-GCM with 128 and 256-bit keys, which were shown to offer the best latency~\cite{tlscore}.
Finally, we evaluate two IPsec tunnel establishment modes: \emph{trap}, where tunnels are created when captured traffic matches the tunnel's policy, and \emph{start}, where tunnels are established immediately at daemon startup.

\begin{figure}[]
    \centering
    \includegraphics[width=0.475\textwidth]{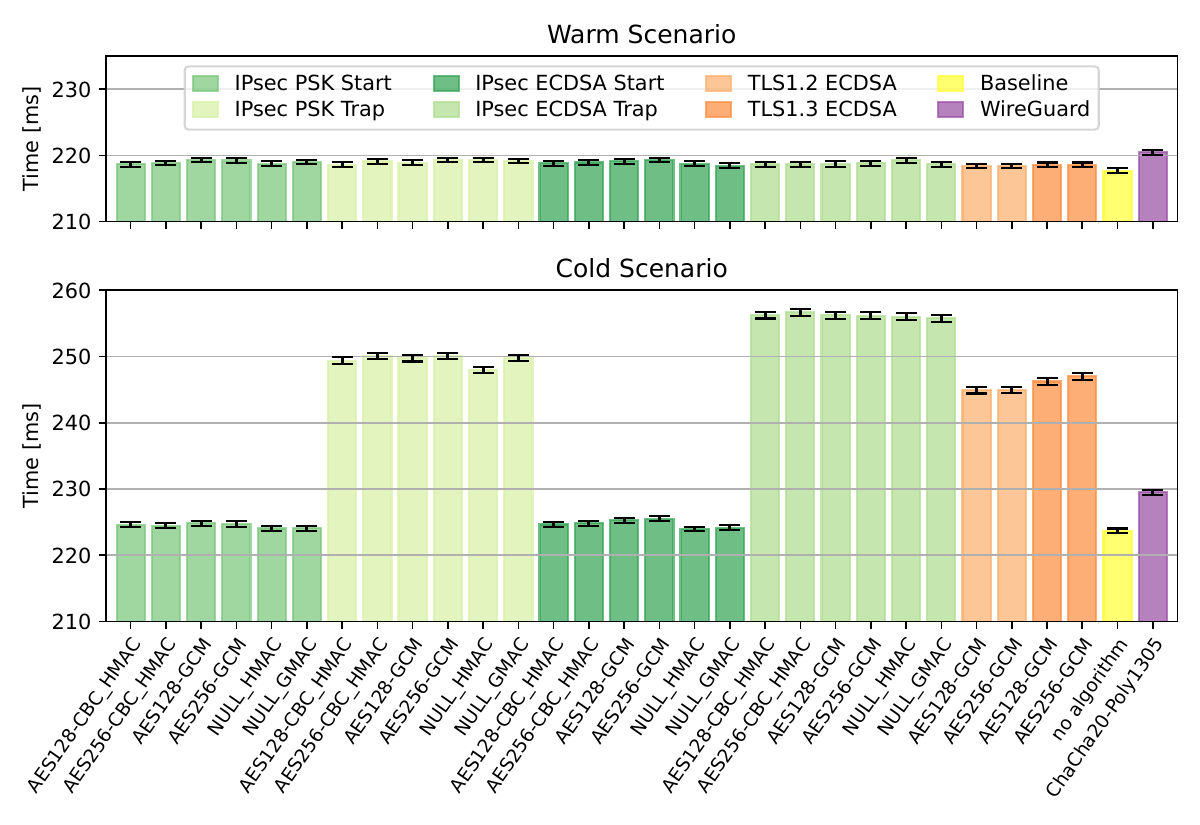}
    \vspace{-20pt}
    \caption{The average UE attachment time in \emph{warm} scenarios is similar across configurations and protocols, except for WireGuard which is slightly slower. In \emph{cold} scenarios, WireGuard and IPsec in start mode demonstrate their advantages.}
    \label{fig:core}
\end{figure}

 \textbf{Results.}
As expected, the results in Figure~\ref{fig:core} show significant differences between the two scenarios. 
In the warm scenario, grouping the configurations of each protocol reveals that TLS is, on average, slightly faster than IPsec, imposing \SI{0.8}{\milli\second} of overhead compared to \SI{1.2}{\milli\second} for IPsec, corresponding to $\sim$0.4\% and 0.55\% increases, respectively.
While TLS is faster on average, multiple IPsec configurations are on par with TLS.
In contrast, WireGuard introduces nearly twice the overhead, making it 1\% slower than the best performing IPsec configuration. %
Overall, all protocols perform very well in this scenario, as the measurements are unaffected by authentication and key exchange, with \ac{ipsec} and \ac{tls} showing a slight edge over WireGuard.
In the cold scenario, where authentication and key exchange delay communication, the results are more scattered. 
A comparison of IPsec authentication methods in \emph{trap mode} shows that \ac{psk}, on average, is faster than certificate-based authentication, due to bypassing certificate verification. 
However, even the fastest IPsec \ac{psk} configuration is slower than the worst-performing \ac{tls} configuration, adding approximately 1 ms more overhead (\SI{23}{\ms} compared to \SI{24}{\ms}). %
WireGuard outperforms both due to its faster handshake, with just \SI{6}{ms} ($\sim$2.5\%) overhead.
However, in \emph{start mode}, where tunnels are established at network startup, IPsec demonstrates a clear advantage over TLS and WireGuard, with configurations adding on average $\pm$\SI{1}{ms} ($\sim$0.4\%). 
This makes IPsec in start mode, on average, \SI{22}{ms} ($\sim$9\%) faster than TLS and \SI{5}{ms} ($\sim$2.15\%) faster than WireGuard.
Across both scenarios, IPsec is the only protocol with latency overhead below 1 ms—approximately \SI{600}{\micro\second} in the warm scenario and \SI{300}{\micro\second} in the cold scenario in the fastest configuration (\emph{ECDSA with NULL\_GMAC}).
On the other hand, while WireGuard is slightly slower than IPsec and TLS in terms of encryption and integrity protection, it remains extremely fast, especially in cold scenarios.
We further discuss the applicability of the protocols in~§\ref{sec:Applicability}.

\subsection{Resource Consumption and Scalability}  
\label{sec:scalability}

\begin{figure}[] 
    \centering
    \includegraphics[width=0.475\textwidth]{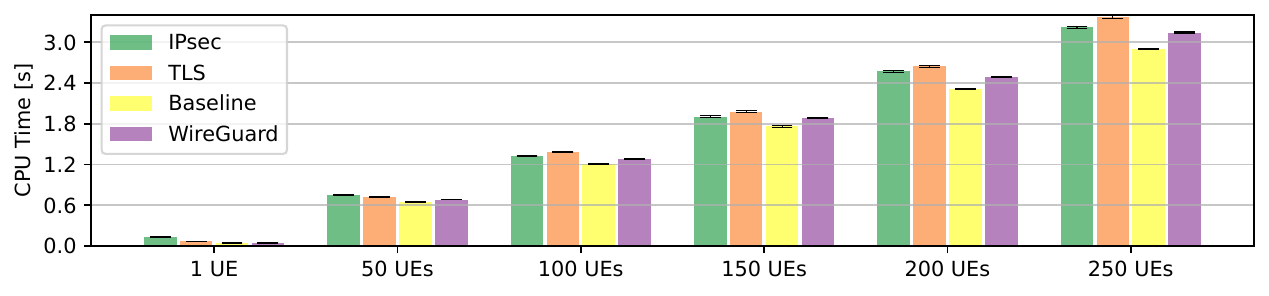} 
    \vspace{-20pt}
    \caption{The average CPU time for the successful registration of increasing UE numbers reveals scales linearly across security protocols, with a slight advantage for WireGuard. }
    \label{fig:scalability}
\end{figure}

Complementing our focus on latency, we also evaluate resource utilization by measuring the CPU time of each protocol, as processes with lower CPU time scale better and are typically more resource-efficient.
To assess scalability, we measure CPU time while scaling the number of UEs, generating a significant amount of traffic within the 5GC CP.
Our measurements track CPU time from boot-up to attachment completion in warm scenarios, repeated 50 times and averaged with 99\% confidence intervals (Fig.~\ref{fig:scalability}). 
Since TLS and IPsec show no measurable difference, we focus on \emph{IPsec with ECDSA and AES-GCM-256 in tunnel mode} and \emph{TLSv1.3 with ECDSA and AES-GCM-256}. 
At larger scales, we expect AEAD schemes (i.e., AES-GCM) to perform better due to their parallelizability (cf. §\ref{label:sec_prot}) and PSK configurations to offer advantages due to the absence of  resource-intensive public-key cryptography.
Our measurements show that all protocols scale linearly, with WireGuard requiring the least resources. 
We expect the gap to widen further in devices without AES acceleration and at larger scales with more UEs.

\subsection{Validation with Real UEs}
\label{sec:realues}
Our measurements are designed to ensure statistical soundness, requiring many samples to minimize the effect of external factors.
To achieve this efficiently, we use simulated UEs, though this may reduce real-world applicability. 
To validate our findings, we extend our testbed to support SRSRAN~\cite{srsran}, which we use with a USRP X310, a Samsung Galaxy A14 5G, and a sysmocom ISIM-SJA5. 
With this more realistic setup, we reproduce selected measurements over N3 and the 5GC, specifically using \emph{WireGuard} and \emph{IPsec in start mode with ECDSA and AES256\_GCM}. 
Across all measurements, latency and jitter increase notably, resulting in wider confidence intervals. 
Specifically, for 5GC measurements, this prevents us from identifying significant differences between the three protocols without additional samples. 
In contrast, over the N3 interface, while IPsec and the baseline remain indistinguishable ($\sim$\SI{29}{\milli\second}), WireGuard consistently shows slightly higher latency ($\sim$\SI{31}{\milli\second})—an increase of approximately 7\%. 
These results support IPsec in start mode as a strong candidate for low-latency communication.

\subsection{Operators' Perspective}
\label{sec:MNOs}
To gain deeper insights into IPsec utilization, the accuracy and impact of our results for real-world 5G deployments, we discuss our findings with two European MNOs, A and B, who request anonymity. 
Operator A confirms using both IPsec and TLS in the 5GC for different use cases. 
While they do not provide exact figures, they prefer TLS in low-latency cases, as internal experiments showed TLS had a lower latency impact on connection establishment than IPsec. 
However, their comparison does not consider IPsec in start mode, which aligns with our findings, suggesting TLS has lower overhead than non-start mode IPsec configurations.
Operator A finds our results promising and plans to investigate IPsec in start mode further. 
On the other hand, no operator currently secures the N3 interface with IPsec. 
Operator B emphasizes the importance of high availability and low complexity in the RAN. 
Unlike 5GC \acp{nf}, which can be redeployed quickly in case of failure, the RAN lacks full virtualization and often requires on-site technicians. 
IPsec introduces challenges, including configuration complexity, certificate management, and reduced traffic visibility. 
Given the RAN's need for continuous operation to avoid legal consequences, e.g., state-imposed fines, the operator deems IPsec an unsuitable option.
In conclusion, while operators are actively exploring security, they sometimes prioritize low complexity, especially where operational stability is critical—making our findings valuable in guiding them to prioritize or balance security, performance, and complexity.

\section{Discussion on Security and Latency}
\label{sec:latencySecurity}

From a security standpoint, both IPsec and WireGuard offer similar guarantees, each with its advantages and drawbacks. WireGuard’s single configuration is resistant to misconfigurations and easy to deploy, but it can be limiting, such as when pre-shared keys are impractical. 
In contrast, IPsec supports a broader range of configurations, providing greater flexibility but also increasing the risk of misconfigurations, a common security concern. 
However, one configuration specifically, \emph{the start mode}, makes IPsec the best configuration in terms of latency, even compared to TLS. 
Our measurements indicate that, in start mode, other configurations have minimal impact on latency. 
Thus, we recommend the most secure setup: \emph{any 256-bit cipher with ECDSA}, as DH-Group 19 ensures perfect forward secrecy, and \emph{tunnel mode}, which encrypts IP headers. 
However, at larger scales or in resource-constrained devices, \emph{PSK configurations with AES-GCM} may offer better latency performance due to their lower resource consumption (cf. Section~\ref{sec:scalability}).

\section{Applicability}
\label{sec:Applicability}
Our results show that both WireGuard and IPsec can meet most application requirements for user data transmission (\emph{over the N3 interface}). 
For time-critical applications, IPsec is the preferred option for UP data transmission due to its slightly better performance in latency. %
However, where simplicity is prioritized, such as in commercial RANs (cf. §\ref{sec:MNOs}), WireGuard is a promising alternative.

\emph{Within the 5GC}, both protocols perform very well, with a slight advantage for IPsec. In commercial deployments where NFs may be frequently deployed (cf. §\ref{sub:control}), IPsec offers faster deployment. 
However, when pre-establishing tunnels (i.e., start mode) is challenging or when simplicity is preferred, WireGuard is the better option.
In static deployments (e.g., in-house industrial 5G) where new NFs are rarely deployed, both protocols are suitable with minimal impact.

Lastly, we expect WireGuard to outperform IPsec on devices without AES acceleration and in resource-constrained environments. 
Our measurements show that while WireGuard's latency performance is slightly behind hardware-accelerated IPsec with AES, it requires fewer resources. 
Its fully parallelizable software implementation of ChaCha20-Poly1305, makes it more resource-efficient across various hardware platforms. 
Additionally, its smaller handshake further reduces network resource consumption, which is particularly beneficial in environments like Narrowband IoT.

\section{Conclusion \& Future Work}
\label{sec:conc}
Our work strives to provide insights into the impact of network-layer security on 5G. To this end, we built a testbed to measure the latency of IPsec and WireGuard within the 5GC CP and during UP data transmission over the N3 interface. Our results show that properly configured IPsec is the best-performing protocol, introducing latency overhead in the microseconds range. WireGuard, while slightly slower, remains a lightweight and efficient alternative.

Future work should examine the impact of security controls on internal RAN and emerging O-RAN interfaces. 
Optimized \ac{ipsec} implementations—e.g., using packet slicing—must be evaluated, as they have shown potential for achieving sub-1\,ms RTTs~\cite{10.1145/3605801.3605818}. 
Additionally, assessing the impact of post-quantum cryptography on 5G and future 6G systems is essential. 
In hybrid scenarios, where authentication and key exchange are more demanding and occur twice, our findings indicate that \ac{ipsec} in start mode may be the only viable option for maintaining ultra-low latency.

\begin{acks}
Funded by the German Federal Office for Information Security (BSI) under project funding reference number 01MO24003B (CSII). The authors are responsible for the content of this publication.
\end{acks}

\end{document}

%% file: acronyms.tex
\DeclareAcronym{urllc}{
    short = URLLC,
    long  =  Ultra-Reliable Low Latency Communication
}

\DeclareAcronym{mmtc}{short= mMTC, long=Massive Machine-Type Communication}

\DeclareAcronym{sbi}{short= SBI, long=Service Based Interface}

\DeclareAcronym{ipsec}{short=IPsec, long= Internet Protocol Security}

\DeclareAcronym{up}{short =UP, long=User Plane}

\DeclareAcronym{cp}{short =CP, long=Control Plane}

\DeclareAcronym{ran}{short =RAN, long=Radio Access Network}

\DeclareAcronym{5gc}{short=5GC, long = 5G Core}

\DeclareAcronym{ue}{short=UE,long=User Equipment}

\DeclareAcronym{nf}{short=NF,long=Network Function}
\DeclareAcronym{gnb}{short=gNB, long=Next Generation Node B}

\DeclareAcronym{upf}{short=UPF,long=User Plane Function}

\DeclareAcronym{amf}{short=AMF,long=Authentication and Mobility Function}

\DeclareAcronym{smf}{short=SMF,long=Session Mangement Function}

\DeclareAcronym{tls}{short=TLS, long = Transport Layer Security}

\DeclareAcronym{ike}{short=IKEv2, long = Internet Key Exchange Version 2}

\DeclareAcronym{esp}{short=ESP,long = Encapsulation Security Payload }

\DeclareAcronym{psk}{short=PSK, long = Pre-Shared Keys}

\DeclareAcronym{nrf}{short=NRF, long = Network Repository Function}

\DeclareAcronym{pcf}{short=PCF, long=Policy Control Function}